\documentclass[%
 aip,
 amsmath,amssymb,
 reprint,%
]{revtex4-1}

\usepackage{graphicx}
\usepackage{amsmath,amssymb}
\usepackage[T1]{fontenc}
\usepackage{lmodern}
\usepackage[utf8]{inputenc}
\usepackage{natbib}
\usepackage{hyperref}
\usepackage{color}
\hypersetup{colorlinks=true,citecolor={blue},linkcolor={blue},urlcolor={blue}}
\usepackage{units}
\usepackage[english]{babel}
\usepackage[normalem]{ulem}
\usepackage{tcolorbox}
\usepackage{amssymb}
\usepackage{stix}
\DeclareUnicodeCharacter{0360}{\eth}

\begin{document}

\title{Directional planar antennae in polariton condensates}

\author{Denis Aristov}
\email{da1u22@soton.ac.uk}
\affiliation{Department of Physics and Astronomy, University of Southampton, Southampton SO17 1BJ, United Kingdom}
\affiliation{Hybrid Photonics Laboratory, Skolkovo Institute of Science and Technology, Territory of Innovation Center Skolkovo, 6 Bolshoy Boulevard 30, building 1, 121205 Moscow, Russia}

\author{Stepan Baryshev}
\affiliation{Hybrid Photonics Laboratory, Skolkovo Institute of Science and Technology, Territory of Innovation Center Skolkovo, 6 Bolshoy Boulevard 30, building 1, 121205 Moscow, Russia}

\author{Julian D. T\"{o}pfer}
\affiliation{Hybrid Photonics Laboratory, Skolkovo Institute of Science and Technology, Territory of Innovation Center Skolkovo, 6 Bolshoy Boulevard 30, building 1, 121205 Moscow, Russia}

\author{Helgi Sigurðsson}
\affiliation{Science Institute, University of Iceland, Dunhagi 3, IS-107, Reykjavik, Iceland}
\affiliation{Institute of Experimental Physics, Faculty of Physics, University of Warsaw, ul.~Pasteura 5, PL-02-093 Warsaw, Poland}

\author{Pavlos G. Lagoudakis}
\affiliation{Hybrid Photonics Laboratory, Skolkovo Institute of Science and Technology, Territory of Innovation Center Skolkovo, 6 Bolshoy Boulevard 30, building 1, 121205 Moscow, Russia}
\affiliation{Department of Physics and Astronomy, University of Southampton, Southampton SO17 1BJ, United Kingdom}

\date{\today}

\begin{abstract}
We report on the realization of all-optical planar microlensing for exciton-polariton condensates in semiconductor microcavities. We utilize spatial light modulators to structure a nonresonant pumping beam into a planoconcave lens-shape focused onto the microcavity plane. When pumped above condensation threshold, the system effectively becomes a directional polariton antenna, generating an intense focused beam of coherent polaritons away from the pump region. The effects of pump intensity, which regulates the interplay between gain and blueshift of polaritons, as well as the geometry of lens-shaped pump are studied and a strategy to optimize the focusing of the condensate is proposed. Our work underpins the feasibility to guide nonlinear light in microcavities using nonresonant excitation schemes, offering perspectives on optically reprogrammable on-chip polariton circuitry.
\end{abstract}

\maketitle

Guiding of light waves in planar structures at the microscale is an important step in development of miniaturized optical technologies, like optical circuits and logic gates~\cite{singh2014all,sasikala2018all}. As a consequence, a variety of different methods of light guiding and focusing were realized in e.g. metamaterials ~\cite{naserpour2018plano,lu2012hyperlenses,khorasaninejad2017metalenses}, surface plasmon-polaritons~\cite{liu2005focusing,kim2008focusing,verslegers2009planar, Shi_OptExpr2017}, phase-change materials~\cite{chen2015engineering}, and photonic crystals~\cite{parimi2003imaging,casse2008nano}. However, a shortcoming of many optical devices is their weak Kerr nonlinear response. Techniques to guide instead highly nonlinear exciton-polariton waves~\cite{Sanvitto_NatPho2011, gao2012polariton,Rozas2020,Niemietz2021,Klaas2019} in the strong light-matter coupling regime could open new possibilities in future light-based circuitry and logic~\cite{Kavokin_NatRevPhys2022}. However, so far, guiding of polariton quantum fluids usually relies on resonant injection techniques or irreversible sample fabrication steps which limits their flexibility in field programmable on-chip technologies~\cite{sanvitto2016road}.

Exciton-polaritons (from here on {\it polaritons}) are bosonic quasiparticles from strongly coupled photonic and excitonic modes in semiconductor microcavities~\cite{kavokin_microcavities_2007}. Polaritons inherit a light effective mass, around $10^{-5}$ of the electron mass, from their photonic component, and strong interactions from their excitonic component. These features permit nonequilibrium Bose-Einstein condensation of polaritons at elevated temperatures~\cite{kasprzak_bose-einstein_2006, Carusotto_RMP2013} and ballistic outflow from localized pumping spots~\cite{Wertz_PRL2012}. Today, structured nonresonant excitation is an established method of inducing localized regions of polariton gain leading to the condensate amplification ~\cite{Wertz_PRL2012}, trapping~\cite{cristofolini2013optical, askitopoulos2013polariton}, vortex manipulation~\cite{dall2014creation}, analogue simulators~\cite{Berloff_NatMat2017}, and artificial lattices~\cite{Pickup_NatComm2020}. Flexibility in excitation control paves the way for creation of optical devices such as polariton transistors~\cite{gao2012polariton, ballarini2013all}, logic gates~\cite{liew2008optical} and interferometers~\cite{sturm2014all}. These practical applications, in conjunction with rapid advances in room-temperature materials~\cite{sanvitto2016road, zasedatelev2019room,Zasedatelev2021}, make polaritons prospective candidates for future technologies based on optical information processing or simulation~\cite{Kavokin_NatRevPhys2022}.  
\begin{figure}
    \centering
    \includegraphics[width = \linewidth]{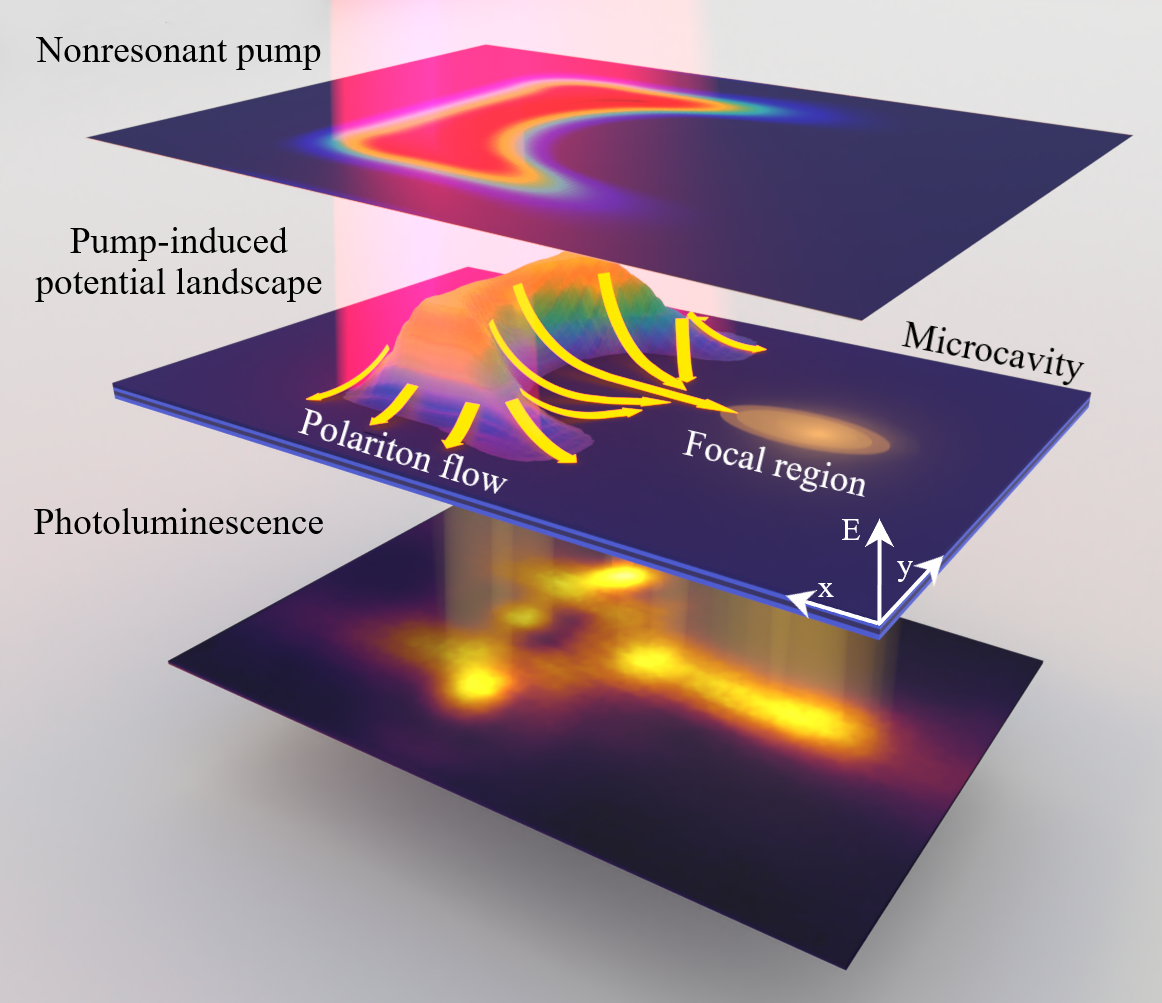}
    \caption{Schematic of an all-optical polariton microlensing effect. Lens-shaped nonresonant pump profile generates a potential landscape for excited polariton waves, which follow the shape of concave lens (radius = 8 $\mu$m, aperture = 16 $\mu$m, thickness = 4 $\mu$m) and focus in the focal region. Yellow arrows in the potential landscape illustrate polariton condensate flow direction. The bottom layer shows experimentally measured cavity photoluminescence from spot 1 corresponding to the condensate density.}
    \label{fig1}
\end{figure}
\begin{figure*}
    \centering
    \includegraphics[width = \linewidth]{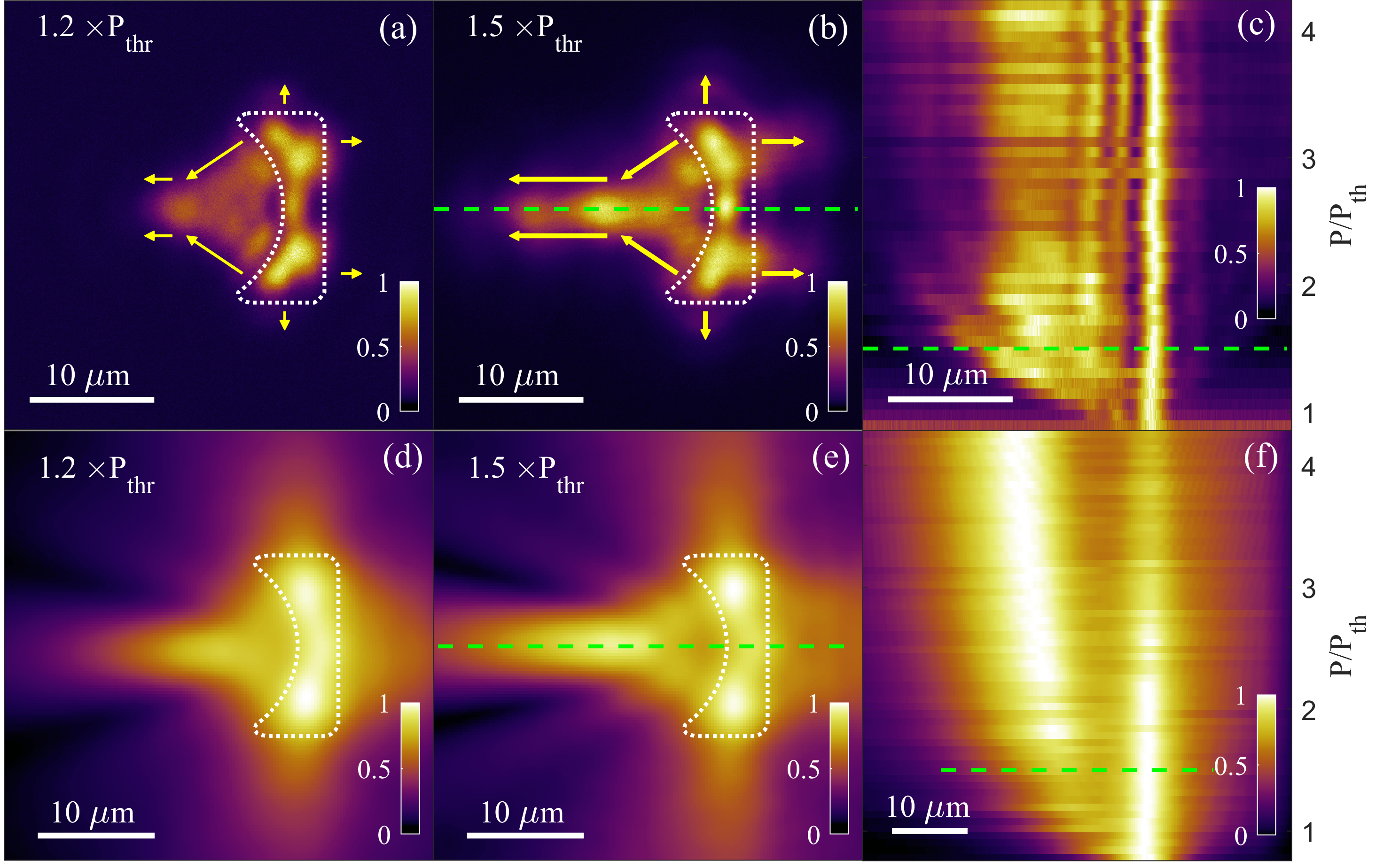}
    \caption{(a,b) Experimental PL from spot 1 for planoconcave lens with radius = 8 $\mu$m, aperture = 16 $\mu$m, thickness = 4 $\mu$m. (c) Line profile along the "optical axis" of the lens (see green dashed line on panel (b)) for varying the pump intensity. Simulated PL for corresponding parameters (d,e) and same line profile along the "optical axis" (green dashed line on panel (e)) of the lens for varying the pump intensity (f). Each panel is normalized independently to increase visibility. White dotted contour represent target for MRAF method for pump profiles generation. Yellow arrows illustrate condensate flow direction.}
    \label{fig2}
\end{figure*}

Due to their large nonlinearities, direct resonant excitation of polaritons was shown as an all-optical method for switching~\cite{Baas_PRA2004, Feng_SciAdv2021, Chen_PRL2022} and to control their planar flow~\cite{Sanvitto_NatPho2011, ballarini2013all, zasedatelev2019room}. But resonant injection demands careful calibration of the excitation beam incident angle and energy which inhibits implementation in integrated on-chip technologies. Instead, nonresonant excitation schemes for controlling the state~\cite{Dreismann_NatMat2016, Klaas_PRB2017} and flow~\cite{gao2012polariton, schmutzler2015all,schmutzler_2016} of condensate polaritons offers a more practical integration into polaritonic devices. Recently, it was proposed that nonresonant excitation beams structured into planar microlenses could act as directional antennas for polariton condensates~\cite{wang2021reservoir}. The reported {\it reservoir optics} scheme exploited the strong interactions and small effective mass of polaritons. In brief, the nonresonant pump photoexcites a co-localized exciton reservoir which in-turn generates and blueshifts polaritons via repulsive polariton-exciton interactions~\cite{Carusotto_RMP2013}. Pumped above condensation threshold, the excited polaritons become macroscopically phase-coherent and thus can interfere constructively when they ballistically flow and refract out of the structured pumping region.

In this letter, we provide an experimental realization of said all-optical plano-concave microlens to guide and focus ballistically propagating condensate polaritons (see Fig.~\ref{fig1}). We employ a strain compensated planar microcavity (two differently detuned spots on the sample are investigated, spot 1 and 2, for more details see supplementary) with embedded InGaAs quantum wells~\cite{cilibrizzi2014polariton}. Sample is held at 4 K and 10 K for spot 1 and spot 2 respectively, and is pumped nonresonantly with a single mode continuous wave laser (see Supplementary Material for experimental parameters). Figures~\ref{fig2}(a) and~\ref{fig2}(b) show the recorded spatially resolved photoluminescence (PL) from spot 1, corresponding to the condensate density, at 1.2$\times P_{th}$ and 1.5$\times P_{th}$, where $P_{th}$ corresponds to the pumping excitation density at condensation threshold. The white dotted lines indicate the boundary of the target region for MRAF, which is effectively the pumped region, and the yellow arrows schematically illustrate the polariton flow. We observe that with increasing excitation density polaritons propagate further away from the excitation area in the direction dictated by the lens shape, implying more efficient focusing [see scan of PL line profiles along the "lens axis" in Fig.~\ref{fig2}(c)]. However, $>1.5 \times P_{th}$ the position and shape of the PL at the focal area starts becoming fixed, indicating a saturation effect.  We note that the in-plane attenuation in the condensate flow is mostly due to the relatively short polariton lifetime, $\approx 5$ ps.

At the lower pumping intensity regime, we observe a nonlinear increase of the condensate's in-plane propagation speed and population with increasing pump intensity just above threshold. The focal distance in this regime and polariton wavevector was predicted to change in proportion with the excitonic reservoir density, which in turn is proportional to the pump intensity~\cite{wang2021reservoir}. This can be observed in a narrow region of pump intensities between 1.1$\times P_{th}$ and 1.5$\times P_{th}$. At the higher pumping intensity regime, i.e. above 1.5$\times P_{th}$, where the reservoir saturates, we observe a slowing down of the change of the effective refractive index so that the position of the brightest focal point remains virtually unaffected of the pump intensity. We point out that the horizontal PL modulations seen in the focal region [see Fig.~\ref{fig2}(c)] can be attributed to weak multi-modal condensation within the pump spot. These results show that the strongest response from the polariton microlens system occurs at lower intensities just above threshold.

 A generalized Gross-Pitaevskii model describes the mean field dynamics of the pumped condensate coupled to an exciton reservoir~\cite{Carusotto_RMP2013} and can be used to qualitatively predict the focusing abilities of reservoir optics elements (see Supplemental Material). The results of numerically solving the coupled nonlinear partial differential equations from a random initial seed recreate the experimentally observed steady state real space PL [Fig.~\ref{fig2}(d,e)] and PL line profiles along the "lens axis" in Fig.~\ref{fig2}(f), showing similar saturation behaviour. 
 
\begin{figure} [h]
    \centering
    \includegraphics[width = \linewidth]{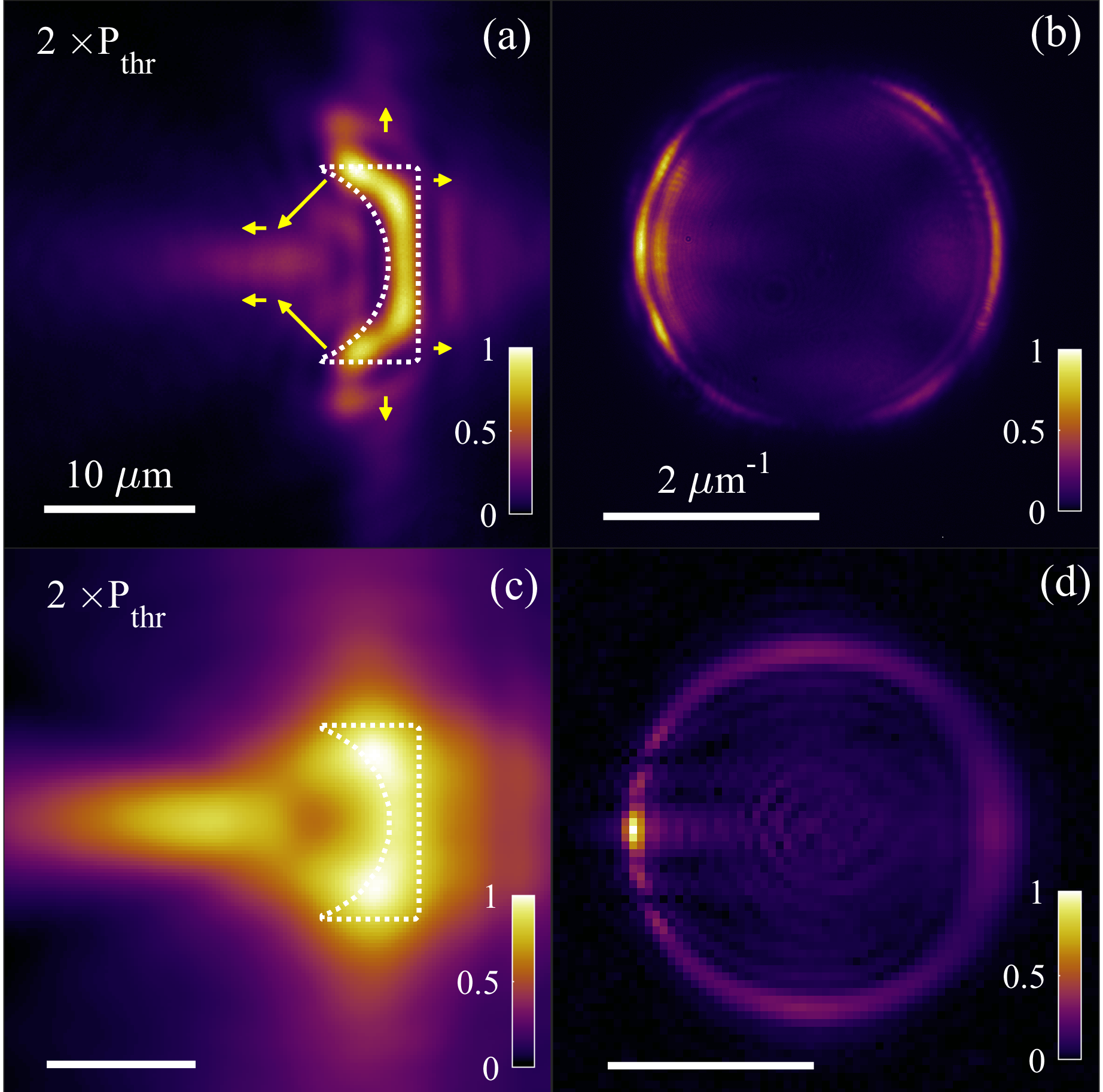}
    \caption{(a) Cavity PL from spot 2 for planoconcave lens with radius = 6.5 $\mu$m, aperture = 13 $\mu$m, thickness = 3 $\mu$m. White outline shows target for MRAF method for pump profile generation. (c) Corresponding simulation result. (b, d)  $k$-space distributions for experiment and theory respectively.) }
    \label{fig2.5}
\end{figure}

These results show that a lens-shaped pump profile creates a polariton steady state condensate wavefunction characterized by a beam of polaritons propagating mostly in one direction (in this instance, the left direction). Latter conclusion can be further strengthened by demonstrating experimental $k$-space distribution [Fig.~\ref{fig2.5}(b)] for planoconcave lens [Fig.~\ref{fig2.5}(a)] and corresponding simulations [Fig.~\ref{fig2.5}(c,d)].

\begin{figure}
    \centering
    \includegraphics[width = \linewidth]{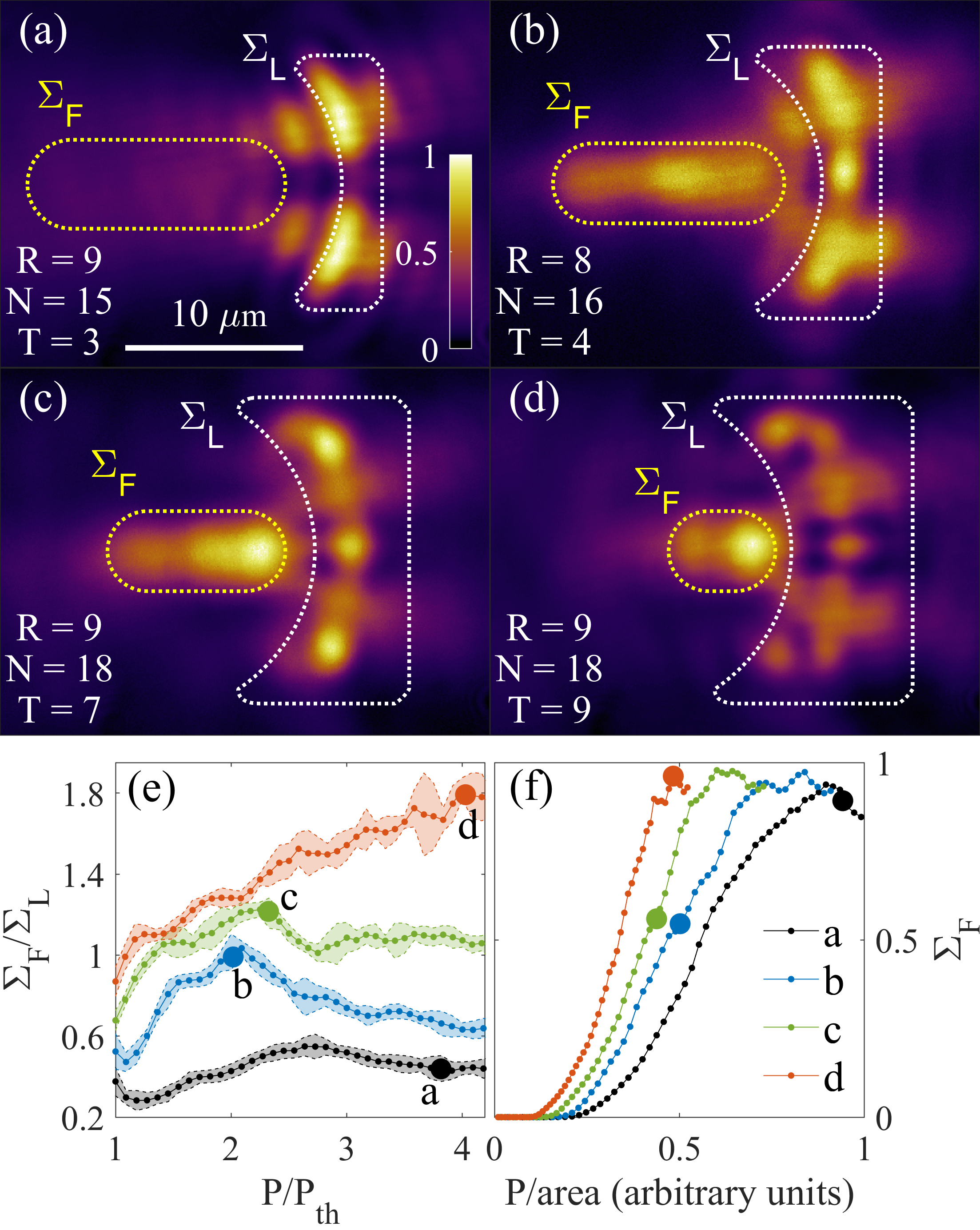}
    \caption{(a-d) Cavity PL from spot 1 for four different lens shapes with curvature radius (R), aperture (N) and thickness (T). White dots outline the lens area and yellow dots the focal area ($\Sigma_{L,F}$ respectively). Each area is chosen ad hoc to lie approximately at the half-maximum contour for pump and PL respectively. Each panel is normalized independently. (e) Corresponding focusing strength $\Sigma_{F}/\Sigma_{L}$ of each lens and (f) and normalized intensities of the focal area $\Sigma_{F}$ as a function of normalized pumping  intensity of each lens shape. The four thick dots on panels (e,f) indicate pump intensities at which PL is shown in panels (a-d).}
    \label{fig3}
\end{figure}

Other reservoir lens parameters, such as aperture (N), thickness (T), and radius of curvature (R), can also be modified freely in our experiment which allows tuning the intensity and propagation distance of guided polaritons. Figure~\ref{fig3}(a) shows a reservoir lens with a curvature radius larger than its half-aperture, resulting in a low condensate fraction in the focal region. Two bright condensate lobes appear within the pump region but are unable to constructively interfere outside. By decreasing the radius of curvature to half of lens aperture and increasing its thickness, we are able to create a highly focused region of polaritons [Fig.~\ref{fig3}(b)]. In general, we observe that lenses of larger thickness, like in Fig.~\ref{fig3}(b-d), result in more localized focal region moving closer to the pumping area. In contrast, thinner lenses like in Fig.~\ref{fig3}(a) result in low focusing. We note that, since the reservoir lens is technically a directional antenna for polaritons, the condensate mode which forms within the pump region plays an essential role in the focusing abilities of the lens. Indeed, we see from all panels in Fig.~\ref{fig3} that complicated "wavefront sources" are being generated within the pump regions which subsequently form complicated refraction patterns, affecting the focusing ability of the lens. More sophisticated pumping geometries can potentially inhibit the different modes forming in the pump region. 

For each case we scan the pump intensity from $1$ to $4$ $\text{P/P}_{\text{th}}$ and plot the \textit{focusing strength} of the lenses [see Fig.~\ref{fig3}(e)], defined as the ratio of average PL intensity in the focal region against the pump region $\Sigma_{F}/\Sigma_{L}$. Here, $\Sigma_{L,F} = \frac{1}{S_{L,F}} \int_{S_{L,F}} I(\mathbf{r}) d\mathbf{r}$ and $S_{L,F}$ corresponds to the areas enclosed by the yellow and white dotted boundaries in Fig.~\ref{fig3}(a-d). We stress that $P_\text{th}$ is different for different lens shapes. Around threshold the PL is mostly emitted from the pump region giving small values of focusing strength. Increasing the pump intensity, we observe how the thickness of the lens dictates its focusing ability with T = 3 $\mu$m lens having smallest focusing strength and lens with T = 9 $\mu$m having highest. One can see that pump intensity curves for (a,b,c) have a point of maximal focusing strength after which the value drops. Since theoretically focusing strength as well as the condensate shape should stay the same after several $P_{th}$ due to the saturation, we explain this effect in experimental results as a consequence of non-ideality in the generated pump profile (see supplementary for more details) and phase-space filling ~\cite{defasing_2014} (i.e., decrease of light-matter coupling). This affects the polariton dispersion stronger in high-pump regime for lenses of small thickness (a,b), less for medium thickness lens (c) and not visible for thickest (d). We also demonstrate the decrease of threshold with increase of lens T by showing the input-output relationship of the average PL in the focal region as a function of pump density in Fig.~\ref{fig3}(f).

We point out that thicker lenses have a larger gain region and therefore a lower pump density threshold for condensation (i.e., "activation") [see Fig.~\ref{fig3}(f) where orange line rises first]. Therefore, the size of the lens can be used to fine-tune the balance between the condensate gain and blueshift coming from the exciton reservoir. As expected, the size of the estimated focusing region becomes smaller since, for standard lenses, it should scale with the wavelength of the wavefront passing through the lens. In general, we observe that thicker lenses pumped at high pump intensities have the strongest focusing abilities as we see in Figs.~\ref{fig3}(d).

Summarizing, we have experimentally demonstrated all-optical and tunable planar microlenses capable of generating and focusing polariton condensate flows up to $25$ $\mu m$ away from the pump region using only nonresonant pumping. Although referred to as \textit{reservoir lenses}~\cite{wang2021reservoir} our pump pattern can also be regarded as optically reprogrammable directional planar antennas for polaritons (i.e., highly nonlinear light). Different lens shapes were presented and analysed, as well as the dependency of the pump intensity on their focusing ability. We showed that tuning of three lens parameters, namely curvature radius, aperture and thickness, yields control over focusing strength and the focal distance. Another advantage of working in the strong light-matter coupling regime is free tuning of the polariton light-matter composition through their photon and exciton Hopfield fractions~\cite{Carusotto_RMP2013}. For this purpose, wedged cavities~\cite{Sermage_PRB2001} offer an additional parameter to tune the focusing ability of our reservoir lenses. Moreover, we have presented results for a relatively short lifetime polaritons $\tau_p \approx  5$ ps which causes strong attenuation as they flow from the pump region. Higher quality cavities with $\tau_p \sim 100$ ps ~\cite{Sun_PRL2017} would allow polaritons to propagate further and possibly offer more focused polariton beams further away from the reservoir lens. 

Polaritonic microlenses offer an advantage in all-optical techniques for polariton manipulation and add to the promising prospect of all-optical polariton computational devices~\cite{Kavokin_NatRevPhys2022}.

See the supplementary material for samples specifications, theoretical model used for simulations, and additional data showing the effect of radius variation on lenses of otherwise fixed geometry. See Supplementary Online Materials for videos showing evolution of condensate over range of increasing pump powers. 

The authors acknowledge the support of the European Union’s Horizon 2020 program, through a FET Open research and innovation action under the grant agreements No. 899141 (PoLLoC) and No. 964770 (TopoLight). H.S. acknowledges the Icelandic Research Fund (Rannis), Grant No. 239552-051.

\section{Author declarations}
\subsection{Conflict of interest}
    The authors have no conflicts to disclose.
\subsection{Author contribution}

$\mathbf{Denis~ Aristov:}$ Conceptualization (equal); Data curation (lead); Formal analysis (lead); Investigation (lead); Methodology (equal); Software (equal); Simulations (lead); Validation (equal); Visualization (lead); Writing - original draft (lead); Writing - review and editing (equal). $\mathbf{Stepan~ Baryshev:}$ Investigation (equal); Project administration (equal); Resources (equal); Writing - review and editing (equal). $\textbf{Helgi~ Sigurðsson:}$ Supervision (equal); Validation (equal); Writing - review and editing (lead). $ \mathbf{Julian ~D.~ T\textbf{\"{o}pfer}:}$ Software (lead); Writing - review and editing (equal). $\mathbf{Pavlos~ Lagoudakis:}$ Conceptualization (lead); Resources (lead);  Supervision (lead);Validation (equal); Writing - review and editing (equal).


\bibliography{references}

\end{document}


\title{Supplementary material: Directional planar antennae in polariton condensates}

\author{Denis Aristov}
\affiliation{Department of Physics and Astronomy, University of Southampton, Southampton SO17 1BJ, United Kingdom}
\affiliation{Hybrid Photonics Laboratory, Skolkovo Institute of Science and Technology, Territory of Innovation Center Skolkovo, 6 Bolshoy Boulevard 30, building 1, 121205 Moscow, Russia}

\author{Stepan Baryshev}
\affiliation{Hybrid Photonics Laboratory, Skolkovo Institute of Science and Technology, Territory of Innovation Center Skolkovo, 6 Bolshoy Boulevard 30, building 1, 121205 Moscow, Russia}

\author{Julian D. T\"{o}pfer}
\affiliation{Hybrid Photonics Laboratory, Skolkovo Institute of Science and Technology, Territory of Innovation Center Skolkovo, 6 Bolshoy Boulevard 30, building 1, 121205 Moscow, Russia}

\author{Helgi Sigurðsson}
\affiliation{Science Institute, University of Iceland, Dunhagi 3, IS-107, Reykjavik, Iceland}
\affiliation{Institute of Experimental Physics, Faculty of Physics, University of Warsaw, ul.~Pasteura 5, PL-02-093 Warsaw, Poland}

\author{Pavlos G. Lagoudakis}
\affiliation{Hybrid Photonics Laboratory, Skolkovo Institute of Science and Technology, Territory of Innovation Center Skolkovo, 6 Bolshoy Boulevard 30, building 1, 121205 Moscow, Russia}
\affiliation{Department of Physics and Astronomy, University of Southampton, Southampton SO17 1BJ, United Kingdom}

\date{\today}
\maketitle

\section{Experimental details} \label{app1}
In this work we used two different spots on a planar, strain compensated $2\lambda$ GaAs microcavities with a Q-factor $ \sim 12000$ and a Rabi splitting $ \sim 8$ meV and negative detuning $ \sim -3$ mEv and $ \sim -2$ mEv correspondingly (spot 1 and spot 2 in main text and further on). Used sample has a following structure: 26 pairs of GaAs and $\text{AlAs}_{0.98}\text{P}_{0.02}$ on bottom and 23 same pairs on top with three pairs of 6 nm $\text{In}_{0.08}\text{Ga}_{0.92}\text{As}$ quantum wells (QWs) embedded at the anti-nodes of the field. Two additional QWs are located at the first and last node of the field to act as a carrier collection wells ~\cite{cilibrizzi2014polariton}.

During the experiments sample was held in cold-finger helium cryostat at 4 K temperature in case of spot 1 and 10 K for spot 2. Nonresonant excitation for all results presented in this work was in the first cavity reflectivity minimum at 783.5 nm. CW laser was chopped by acousto-optic modulator with frequency of pulses 1 kHz and $10\%$ duty cycle to avoid sample overheating. For generation of lens-shaped pump profile initial beam was modulated using phase-only spatial light modulator (SLM). After modulation and isolation of zero-order reflection beam is focused under normal incidence on the sample using microscope objective with NA $= 0.42$. Real-space images of lens were made using transmission PL for spot 1 and reflection PL for spot 2 with same model ORCA cameras.

Phase map for the SLM for each specific lens configuration was calculated using mixed region amplitude freedom (MRAF) method. Presented in the main text figures white outlines of pump areas are used as a targets for MRAF method and since it is approximation technique, resulting pump profiles do not share ideally flat surface of targets. One can see experimental pump profiles in Fig.~\ref{figS1}(g-l). This is the reason, why experimental PL in all figures has nonuniform, yet symmetric, density distribution. This non-ideality of lens profile, indeed, limits performance of such planar element, especially when it comes to finding appropriate starting phase generation parameters for MRAF method, since at high powers even small divergence of pump profile from symmetric with respect to optical axis causes severe change in effective refractive index of the lens, which result in formation of deformed condensate with no dominant flow direction.

\section{Theoretical model} \label{app2}
In this work two-dimensional polariton condensate macroscopic wavefunction $\psi(\mathbf{r},t)$ is modeled using the well known generalized Gross-Pitaevskii equation coupled to a rate equation describing the laser-driven exciton reservoir $X(\mathbf{r},t)$ ~\cite{Carusotto_RMP2013},
\begin{align} \notag
    & i\frac{\partial \psi}{\partial t}  = \bigg[-\frac{\hbar\nabla^{2}}{2m} + \frac{i}{2}\left(R X - \gamma\right) + \alpha |\psi|^{2} \\ 
    & + G \left(X + \frac{P(\mathbf{r})}{W} \right) \bigg]\psi ,
    \label{eq.GPE},
    \tag{S1}\\
    & \frac{\partial X}{\partial t}  = -\left(\Gamma  + R|\psi|^2\right)X + P(\mathbf{r}).
    \label{eq.Res}
    \tag{S2}
\end{align}

Here, $m$ is the polariton mass, $\gamma^{-1}$ the polariton lifetime, $G = 2 g |\chi|^2$ and $\alpha = g |\chi|^4$ are polariton-reservoir and polariton-polariton interaction strengths, respectively, $g$ is the exciton-exciton Coulomb interaction strength, $|\chi|^2$ is the excitonic Hopfield fraction of the polariton, $R$ is the scattering rate of reservoir excitons into the condensate, $\Gamma$ is the reservoir decay rate, $W$ quantifies additional blueshift coming from a background of high-energy and dark excitons generated by the nonresonant continuous-wave pump $\mathbf{P}(\mathbf{r}) $ is pump profile. The parameters are based typical GaAs microcavity properties and fitting to previous experiments ~\cite{Topfer_ComPhys2020} and theoretical investigations ~\cite{wang2021reservoir}: $m = 5 \times 10^{-5} m_0$ where $m_0$ is the free electron mass; $\gamma^{-1} = 2.5$ ps; $|\chi|^{2}=0.4$ since our cavity is negatively detuned; $\hbar g = 1 \, \mu\mathrm{eV\,\mu m^{2}}$; $R=2.8g$; and $W = \Gamma = \gamma$. 

In order to recreate particular distribution of measured from microcavity photoluminescence, pump profile was shaped into most close to the real one form. At first, "ideal" lens-shaped pump profile was formed. Such profile, as discussed in ~\cite{wang2021reservoir}, can be created by applying Gaussian blurring to lens-shaped Heaviside function $F(\mathbf{r})$:
\begin{equation} \label{eq.gaus_blur}
    f(\mathbf{r}) = \frac{1}{2\pi\omega^2}\int F(\mathbf{r'}) e^{\frac{-|\mathbf{r}-\mathbf{r'}|^2}{2\omega^2}}d\mathbf{r'},
    \tag{S3}
\end{equation}
where
\begin{equation} \label{eq.step_lens}
    F(\mathbf{r}) = \begin{cases} 1 & \mathbf{r} \in \emph{L} \\
                     0 & \mathbf{r} \not\in \emph{L}
       \end{cases}
       \tag{S4}
\end{equation}

and \emph{L} is the lens area. Gaussian blurring is a decent approximation, accounting for finite resolution of SLMs and exciton diffusion from the excitation spot. One can see, that for reasonable value of $\omega \approx 0.85$ $\mu m$ ~\cite{wang2021reservoir}, corresponding to $2$ $\mu m$ full-width at half-maximum, areas of the lens more than 2 $\mu m$ away from it's border are slightly affected by the blurring and form a uniform and smooth plato. Such pump profile with plato in the center and smoothly blurring edges is an ideal scenario which in reality cannot occur. That's why $F(\mathbf{r})$ is generated not by simple ~\eqref{eq.step_lens}, but with profile, generated by MRAF algorithm with initial gaussian profile and target, having the form  \emph{L}. After finite number of steps, which should not be too high in order to avoid overfitting and account for finite pixel size of real SLM, resulting $~{F(\mathbf{r})}$ is blurred and applied as a nonresonant pump in simulation.

\section{Additional data on radius variation}
\begin{figure*} [h]
    \renewcommand{\figurename}{Figure}
    \renewcommand{\thefigure}{S\arabic{figure}}
    \centering
    \includegraphics[width = \linewidth]{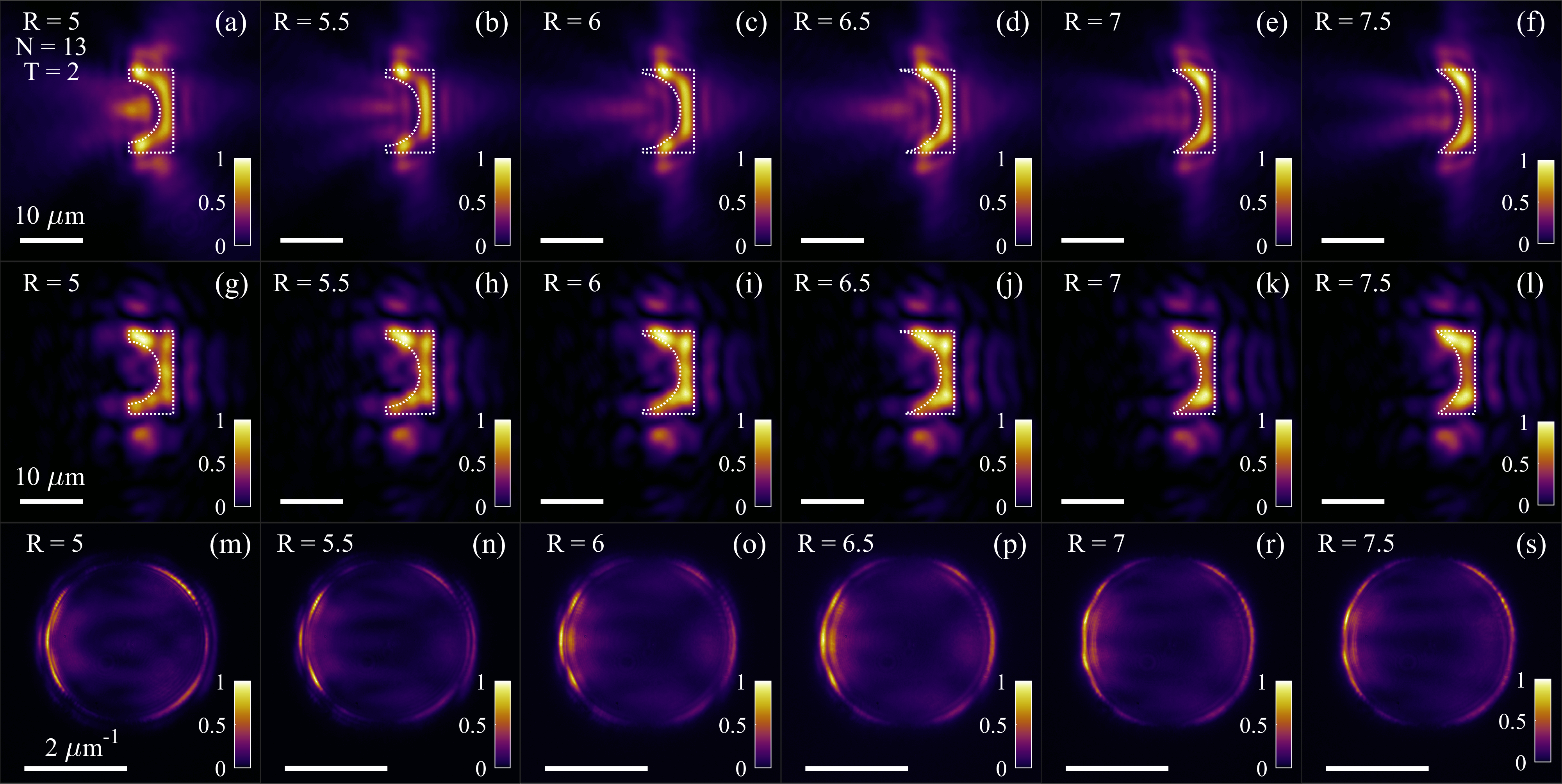}
    \caption{Experimental PL from spot 2 for planoconcave lens shaped pump profile for fixed aperture N = 13 $\mu$m and thickness T = 2 $\mu$m and varying curvature radius (a-f). Experimentally recorded pump profile for corresponding geometries (g-l) and k-space distribution (m-s). Each panel is normalized independently to increase visibility. White dotted lines represent target pump profiles used for MRAF algorithm.}
    \label{figS1}
\end{figure*}

 In Fig.~\ref{figS1} experimental PL for lens profile with all geometrical parameters except for curvature radius is presented, as well as experimental pump profiles and condensate k-space distributions. Pump power in all shots is approximately 3.5$\times P_{th}$. 

\section{Pump power dependence} \label{app2}
In Supplementary Online Materials one can find two multimedia files, showing the evolution of condensate PL for lens shapes with parameters radius (R), aperture (N) and thickness (T) corresponding to panels in the main text with R = 8 $\mu$m, N = 16 $\mu$m, T = 4 $\mu$m  and  R = 9 $\mu$m, N = 18 $\mu$m, T = 7 $\mu$m  over different pump powers.


\bibliography{references}